\pdfoutput=1
\RequirePackage{ifpdf}
\ifpdf 
\documentclass[pdftex]{sigma}
\else
\documentclass{sigma}
\fi

\numberwithin{equation}{section}

\newtheorem{Theorem}{Theorem}[section]
\newtheorem{Proposition}[Theorem]{Proposition}
 { \theoremstyle{definition}
\newtheorem{Definition}[Theorem]{Definition}
\newtheorem{Example}[Theorem]{Example} }

\def\C{\mathbb{C}}
\def\a{{a}}
\def\b{{b}}

\begin{document}
\allowdisplaybreaks

\newcommand{\arXivNumber}{2009.09854}

\renewcommand{\PaperNumber}{067}

\FirstPageHeading

\ShortArticleName{A New Class of Integrable Maps of the Plane}

\ArticleName{A New Class of Integrable Maps of the Plane:\\ Manin Transformations with Involution Curves}

\Author{Peter H.~VAN DER KAMP}

\AuthorNameForHeading{P.H.~van der Kamp}

\Address{Department of Mathematics and Statistics, La Trobe University, Victoria 3086, Australia}
\Email{\href{mailto:P.vanderKamp@LaTrobe.edu.au}{P.vanderKamp@LaTrobe.edu.au}}
\URLaddress{\url{https://scholars.latrobe.edu.au/pvanderkamp}}

\ArticleDates{Received January 15, 2021, in final form July 02, 2021; Published online July 13, 2021}

\Abstract{For cubic pencils we define the notion of an involution curve. This is a curve which intersects each curve of the pencil in exactly one non-base point of the pencil. Involution curves can be used to construct integrable maps of the plane which leave invariant a~cubic pencil.}

\Keywords{integrable map of the plane; Manin transformation; Bertini involution; invariant; pencil of cubic curves}

\Classification{14E05; 14H70; 37J70; 37K60}

\section{Introduction}
One clear definition of an integrable map is the notion of Liouville integrability, which requires the existence of sufficiently many invariant functions in involution with each other \cite{HJN,Ves}. One way to obtain such maps is by reduction from integrable lattice equations \cite{PNC,KQ}. For~pla\-nar maps, integrability is equivalent to the preservation of a pencil of curves and measure-preservation~\cite{RQ}. A rather large (18-parameter) family of integrable maps of the plane was obtained by Quispel, Roberts and Thompson (QRT)~\cite{QRT,QRT-II}. These maps preserve a pencil of biquadratic curves, and have been studied in the context of algebraic geometry of elliptic surfaces in \cite{Dui,IR,Tsu}. It follows from \cite{JRV} that birational maps preserving a pencil of curves are necessarily the composition of two involutions. The classification of birational involutions, of $\mathbb{P}^2$, is a classical problem \cite{BB,Ber} and has lead to the following three types: De Jonquieres involutions, Geiser involutions, and Bertini involutions.

Consider a linear pencil of cubic curves in the $(u,v)$-plane of the form
\begin{equation}\label{P}
P(C):= F(u,v) - C G(u,v) = 0,\qquad C\in P_1(\C).
\end{equation}
Such a pencil has (at most) $3^2$ base points, which are the solutions of $F=G=0$ (Bezout's theorem). A straight line through a (non-singular) base point $p$ will intersect each curve in the pencil in 2 other points. Thus one can define a map which interchanges these 2 points. Such a~map, denoted $\iota_p$, is coined a Manin \cite{Man} involution in \cite[Section~4.2]{Dui} and a $p$-switch in \cite{KMQ}. The composition of two Manin involutions $\tau_{p,q}=\iota_q\circ\iota_p$ is an integrable map of the plane, as it leaves invariant a pencil of curves and is measure preserving \cite[Proposition~8]{KMQ}.

Similarly, one can construct involutions that leave invariant pencils of curves of degree $N=2$, or $N=4$ \cite{KMQ}. For $N=2$, the point $p$, which is called the involution point of $\iota_p$, can be chosen to be any non-base point. For $N=4$, one requires the pencil to have two base points which are double points of both $F=0$ and $G=0$, and these points are taken as involution points. In~\cite{KMQ} it was shown that this construction does not generalise to maps which preserve a pencil of degree $>4$ and that all generalised Manin transformations obtained in this way are equivalent to a QRT map \cite{QRT,QRT-II} through a projective collineation. Recall that the QRT map preserves a~special $N=4$ pencil, namely a biquadratic pencil. In $\mathbb{P}^2$ such a pencil has 2 double base points, at $(1:0:0)$ and $(0:1:0)$. The corresponding involutions are the horizontal switch $\iota_1$, and the vertical switch $\iota_2$, and the QRT map is the composition $\tau=\iota_2 \circ \iota_1$ \cite[p.~viii]{Dui}.

In \cite{PSWZ} a new type of Manin involutions was introduced. Given a pencil $P$, another pencil~$V$ is constructed such that each curve in $P$ intersects each curve in~$V$ in two non-base points, which are interchanged by the involution. This construction includes $\iota_p$ as a special case, where~$V$ has degree $D=1$ and base point~$p$, cf.\ Figure~\ref{f0}. In general, the pencil~$V$ can have a degree $D>1$. For example, consider a pencil~$P$ of degree $N=4$ with 2 double base points and 8 simple ones. Taking 4 base points of $P$, including the~2 double base points, one constructs the unique pencil~$V$ of degree $D=2$ which has those points as simple base points. Then each curve of~$V$ intersects each curve of~$P$ in precisely two other points. Thus an involution can be defined that switches these 2 points.

In this paper, we describe another type of Manin involutions. Recently, a planar map $\gamma$ \cite[equation~(1.1)]{CKZ} was obtained by taking an open boundary reduction from the $Q1(\delta=0)$ quad-equation, and it was shown to be a composition of two involutions, $\gamma=\iota_q\circ \iota_p$,
where one involution point, e.g.~$p$, is a simple base point of a singular cubic pencil, and the other involution point, $q=q(C)$, depends on the particular curve in the pencil. Such involution points can be obtained as the unique non-base point intersection of a so-called involution curve with the pencil. The Bertini involution can be understood as a Manin involution with an involution curve of degree 4 with a triple base point at one of the base points of the invariant cubic pencil, cf.~\cite{Moo}. The birational equivalence of other Manin involutions with involution curves to De Jonquieres involutions or Geiser involutions will not be studied here. As the reader may have noticed, the term `Manin involution' is used to denote a variety of involutions, including ones that leave invariant quadratic pencils, and degenerate pencils of singular cubics, which could be called degenerate Manin involutions. We also note that a $p$-switch in one set of coordinates may not be a $p$-switch after a birational transformation, i.e., when the transformation is not a collineation. As usual, a non-trivial composition of two Manin involutions is called a Manin transformation. If one of the involutions, or both, is a Manin involution with an involution curve, the composition is called a Manin transformation with an involution curve. We describe several geometric settings in which involution curves lead to classes of maps that leave invariant pencils of cubic curves and are measure-preserving. We show that
for singular cubic pencils there are two classes of involution curves of degree $M$, for any $M$. For non-singular cubic pencils the number of classes of involution curves seems to grow linearly with the degree.

\section[A map obtained by open reduction of the Q1(delta=0) equation]
{A map obtained by open reduction of the $\boldsymbol{Q1(\delta=0)}$ equation}

The map $\gamma$ found in \cite{CKZ} is given by
\begin{gather*} 
\gamma(u,v)
=\frac{(u+v)(\a u + \b(\a u + v+1)v)^{2}}{\b(u+(u\b+v+1)v)Z}
\bigg(\frac{u (\a u+ (\a\b u + \b v +\a)v)}{\a (u+(\a u+v+1)v)},v\bigg),
\end{gather*}
where $Z=
\big(\a^{2}+\b\big)u{v}^{2}+\a\big(\b{u}^{2}+{v}^{2}\big)v+\a(u+v)^2$. It leaves invariant a singular pencil of cubic curves (of genus~0), of the form~(\ref{P}) with
\begin{gather} \label{FaG}
F(u,v)=v^2 (1 + u + v) + \a u \bigg(\frac{u}{\b} + u v + v^2\bigg),\qquad
G(u,v)=uv.
\end{gather}
This pencil has 5 base points in homogeneous coordinates
\begin{gather*}
b_0=(0:0:1),\qquad
b_1=(0:-1:1),\qquad
b_2=(1:0:0),
\\
b_3=(1:-1:0),\qquad
b_4=(1:-\a:0),
\end{gather*}
of which the first one is a double point. The simple base points yield the following $b$-switches
\begin{gather*}
\iota_{b_1}(u,v)=
\bigg(\frac {u \big(ab{u}^{2}v+abu{v}^{2}+abuv+bu{v}^{2}+b{v}^{3}+a{u}^{2}+buv+2 b{v}^{2}+vb\big)}
{bv(v+1) (v+u+1) (au+v+1)},
 \\ \hphantom{\iota_{b_1}(u,v)=\bigg(}
\frac{a{u}^{2}}{(au+v+1)(v+u+1)vb}\bigg),
 \\
\iota_{b_2}(u,v)=\bigg(\frac {b{v}^{2} (v+1)}{ua (vb+1)},v\bigg),
\\
\iota_{b_3}(u,v)=\bigg(\frac{vb (v+u+1)(u+v)}{buv+b{v}^{2}+au+vb},
\frac{au (u+v)}{buv+b{v}^{2}+au+vb}\bigg),
\\
\iota_{b_4}(u,v)=\bigg(\frac {vb(au+v+1)(au+v)}{a\big(abuv+b{v}^{2}+vb+u\big)},
\frac {u(au+v)}{abuv+b{v}^{2}+vb+u}\bigg).
\end{gather*}
One can check that each of the compositions $\gamma\circ \iota_{b_i}$ and $\iota_{b_i}\circ\gamma$ is a involution. In fact, they are $p$-switches, where $p$ depends not only on $a$, $b$ but also on $C$, the parameter of the pencil $P(C)$. We define $\gamma\circ \iota_{b_i}=\iota_{h_i}$ and $\iota_{b_j}\circ\gamma=\iota_{k_j}$.

The involution point of a $p$-switch $\delta$, which leaves invariant the same pencil as $\gamma$, can be calculated as follows. Starting with $x_1=(u,v)$, determine
\[
x_2=\delta(x_1),\qquad
x_3=\gamma(x_1),\qquad
x_4=\delta(x_3).
\]
The involution point $p$ of $\delta=\iota_p$ is obtained as the intersection of the lines $x_1x_2$ and $x_3x_4$.

Following this procedure, for each map $\iota_{h_i}$ and $\iota_{k_j}$, the variables $u$,~$v$ can be eliminated, and explicit expressions for the involution points $h_i$ and $k_j$ in terms of $a$, $b$, $C$ can be obtained, cf.~\cite{CKZ}, where $h_1$ and $h_2$ were provided, and Appendix~\ref{appendixA}. The points $h_i$, $k_j$ are in the intersection of the pencil $P(C)$ with curves that we denote by $H_i$, $K_j$. These curves can be obtained as follows. For an involution point $p(a,b,C)$ eliminate (using a Groebner basis) the variable $C$ from the set of equations $\{u=p_1,v=p_2\}$. We obtain
\begin{gather*}
H_1:=uvab+{v}^{2}b+au+av=0,
\\
H_2:=auv+{v}^{2}+u+v=0,
\\
H_3:=vb{u}^{2}a+a{u}^{2}+ \big({a}^{2}+b\big) u{v}^{2}+2 auv+a{v}^{3}+a{v}^{2}=0,
\\
H_4:=uvb+{v}^{2}+u+v=0,
\\
K_1:={a}^{2}b{u}^{2}v+{a}^{2}{u}^{2}+ \big(b{a}^{2}+ab\big) u{v}^{2}+2
 uvab+b{v}^{3}a+{b}^{2}{v}^{2}=0,
\\
K_2:={u}^{2}va{b}^{2}+{a}^{2}{u}^{2}+ \big(a{b}^{2}+{b}^{2}\big) u{v}^{2}+2 uvab+{b}^{2}{v}^{3}+{b}^{2}{v}^{2}=0,
\\
K_3:=au+vb=0,
\\
K_4:={a}^{2}b{u}^{2}v+{a}^{2}{u}^{2}+ \big(b{a}^{2}+{b}^{2}\big) u{v}^
{2}+2 uvab+{b}^{2}{v}^{3}+{b}^{2}{v}^{2}=0.
\end{gather*}
Each of the above curves intersects the pencil $P(C)$, with $F$, $G$ given by (\ref{FaG}) in some (or all) of~the base points of the pencil. The type of intersection at the base points of the pencil is given in Table~\ref{BPI}. Note that all curves in $P(C)$ are tangent to each other at base point $b_2$, i.e., there are two infinitely near base points, $b_2$ and $b_2^\prime$.
\begin{table}\centering\renewcommand{\arraystretch}{1.2}\setlength{\tabcolsep}{6.5pt}
\caption{Degrees of the curves defined by involution points, and their multiplicities.}
\vspace{1ex}\label{BPI}
\begin{tabular}{c|c|c|c|c|c|c|c}
\hline
Curve & Degree & $b_0$ & $b_1$ & $b_2$ & $b_2^\prime$& $b_3$ & $b_4$\\
\hline
$H_1$ & $2$ & $1$ & $0$ & $1$ & $1$ & $0$ & $1$ \\
$H_2$ & $2$ & $1$ & $1$ & $1$ & $0$ & $0$ & $1$ \\
$H_3$ & $3$ & $2$ & $1$ & $1$ & $1$ & $0$ & $1$ \\
$H_4$ & $2$ & $1$ & $1$ & $1$ & $1$ & $0$ & $0$ \\
$K_1$ & $3$ & $2$ & $0$ & $1$ & $1$ & $1$ & $1$ \\
$K_2$ & $3$ & $2$ & $1$ & $1$ & $0$ & $1$ & $1$ \\
$K_3$ & $1$ & $1$ & $0$ & $0$ & $0$ & $0$ & $0$ \\
$K_4$ & $3$ & $2$ & $1$ & $1$ & $1$ & $1$ & $0$ \\
\hline
\end{tabular}
\end{table}
The intersection number of the quadratic curve $H_1$ with the pencil $P(C)$ equals $2\cdot3=6$. On the other hand, the sum of the intersection numbers at the base points is $2+0+(1+1)+0+1=5$. This means that there is only one non-base point in the intersection of the two curves. Similarly, the intersection number of the cubic curve~$K_2$ with the pencil $P(C)$ equals $3\cdot3=9$, and the sum of the intersection numbers at the base points is $4+1+(1+0)+1+1=8$. In fact, one can check that for all cur\-ves~$H_i$,~$K_i$, apart from the intersection at the base points, there is only one non-base point simple intersection between the curve and the pencil. This makes it possible to define, for each curve in the pencil, the involution point to be this unique non-base point intersection.

A further simplification of the map $\gamma$ is possible. The pencil $P(C)$, with (\ref{FaG}), is a special subset (i.e., those curves which contain $b_1^\prime$) of the net of cubic curves with a zero of the second order at $b_0$ and zeros of the first order at $b_1$, $b_2$, $b_3$, and $b_4$,
\[
N=\big\{c_1 F(u,v) + c_2 G(u,v) + c_3 H(u,v) = 0 \mid (c_1:c_2:c_3)\in P_2(\C)\big\},
\]
with $H(u,v)=u^2$. By change of variables $x=F(u, v)/G(u, v)$, $y=H(u,v)/G(u, v)$ the map $\gamma$ is transformed into
\begin{gather} \label{gxy}
\gamma(x, y)=\bigg(x,-\frac{b((2a + bx)y + a- b)}{a((a -b)y -b(x + 2))}\bigg),
\end{gather}
which is the case of a birational map preserving the structure of a
ruled surface in the classification by \cite{DF}. In terms of $x$,~$y$, each involution $\iota_{b_i}$, $\iota_{h_i}$ and $\iota_{k_i}$ has the same form as (\ref{gxy}), leaving~$x$ invariant and the image of $y$ being fractional linear in $y$, and we still have
\begin{gather} \label{decomp}
\gamma=\iota_{h_i}\circ \iota_{b_i}=\iota_{b_i}\circ\iota_{k_i},\qquad i=1,\ldots,4.
\end{gather}
The algebraically simplest representation is $\gamma=\iota_{h_4}\circ \iota_{b_4}$ with
\[
\iota_{h_4}(x,y)=\bigg(x,\frac{b(bx-ay+a)}{a(ay+b)}\bigg),\qquad
\iota_{b_4}(x,y)=\bigg(x,\frac{bx-ay+b}{a(y+1)}\bigg).
\]
We note that the decompositions (\ref{decomp}) of $\gamma(x,y)$ would be hard to find if you were only given~(\ref{gxy}), as all base points have disappeared. Moreover, in terms of $x$,~$y$ the involutions are no longer $p$-switches; each involution leaves invariant, besides $x$, one or more multi-parameter families of~curves quadratic in $y$.

\section{Involutions curves for cubic pencils}

In the sequel, we refer to a curve which consists of involution points as an involution curve.
\begin{Definition}
A curve $Q$ is an {\it involution curve} for a cubic pencil $P(C)$ if the intersection numbers of $Q$ with any curve of $P(C)$ at the base points of $P(C)$ add up to $3\deg(Q)-1$.
\end{Definition}

Let $Q$ be an involution curve for a cubic pencil $P(C)$ whose set of base points is denoted $B$. An involution $\iota_Q$ can be constructed as follows. For any given $p\not\in B$ let $C_p$ be the value of $C$ such that $p$ is on the curve $R=P(C_p)$. We take $q$ to be the unique non-base point intersection of $Q$ and $R$, and define the restriction of $\iota_Q$ to $R$ to be simply given by the $q$-switch $\iota_q$. The following four cases belong to the standard definition of a $q$-switch. If $p=q$ is a flex point of $R$ then $\iota_Q(p)=p$. If $p=q$ is a not a flex point of $R$ then $\iota_Q(p)$ is the unique intersection $r\neq p$ of the tangent line to $R$ at $p$ and $R$. If $p\neq q$ and the line $pq$ is tangent to $R$ at $p$ then $\iota_Q(p)=p$. Finally, if $p\neq q$ intersects $R$ in a third point $r$, then $\iota_Q(p)=r$.

\begin{figure}[h]\centering
\includegraphics[width=11cm]{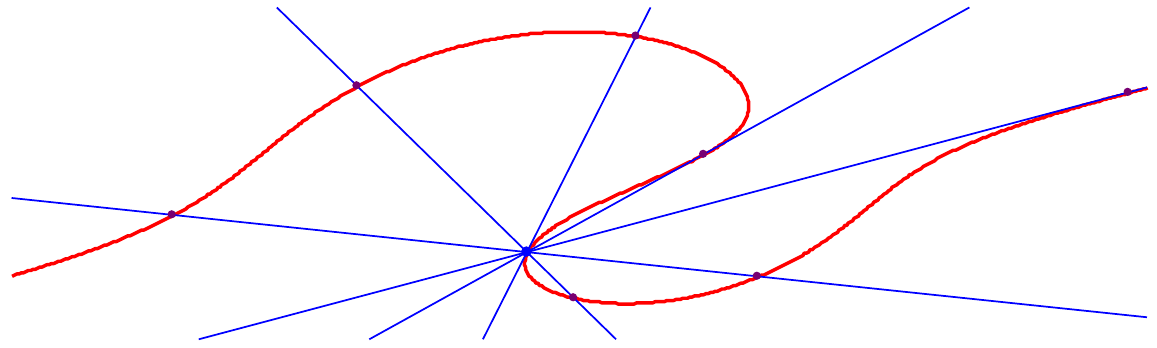}
\caption{The action of a $q$-switch on a cubic curve. Three of the straight blue lines are tangent to~the red curve $R$, and for two points $p$ we have $\iota_q(p)=p$. The five blue lines are members of a pencil of~degree $M=1$, whose base point is the blue dot.}\label{f0}
\end{figure}

In Figure~\ref{f0} we illustrate the action of a $q$-switch on a cubic curve $R$ by connecting points $p\in R$ to their images $\iota_q(p)$ using straight lines through the involution point $q\in R$.

\subsection{Involution curves for singular cubic pencils}
The pencil in the previous section, apart from being singular, has a special extra feature, namely that all curves are tangent in one point. For such pencils the following is clear.
\begin{Proposition} \label{P1}
Let $P(C)$ be a pencil of cubic curves, which has $1$ double base point $b_0$, two infinitely near base points $b_1$, $b_1^\prime$ $($so each curve in $P(C)$ has the same tangent line at $b_1$, say $T)$, and simple base points $b_2$, $b_3$, $b_4$. Let $Q$ be a curve such that either
\begin{itemize}\itemsep=0pt
\item $\deg(Q)=2$, $b_0$ is on $Q$, $b_1$, $b_1^\prime$ are on $Q$, and $Q$ contains one other point $b_i$ with $i\in\{2,3,4\}$, or
\item $\deg(Q)=3$, $b_0$ is a double point on $Q$, $b_1$, $b_1^\prime$ are on $Q$,
 and $Q$ contains two distinct points~$b_i$ with $i\in\{2,3,4\}$.
\end{itemize}
Then $Q$ is an involution curve.
\end{Proposition}

\begin{figure}[h]\centering
\includegraphics[width=6.5cm]{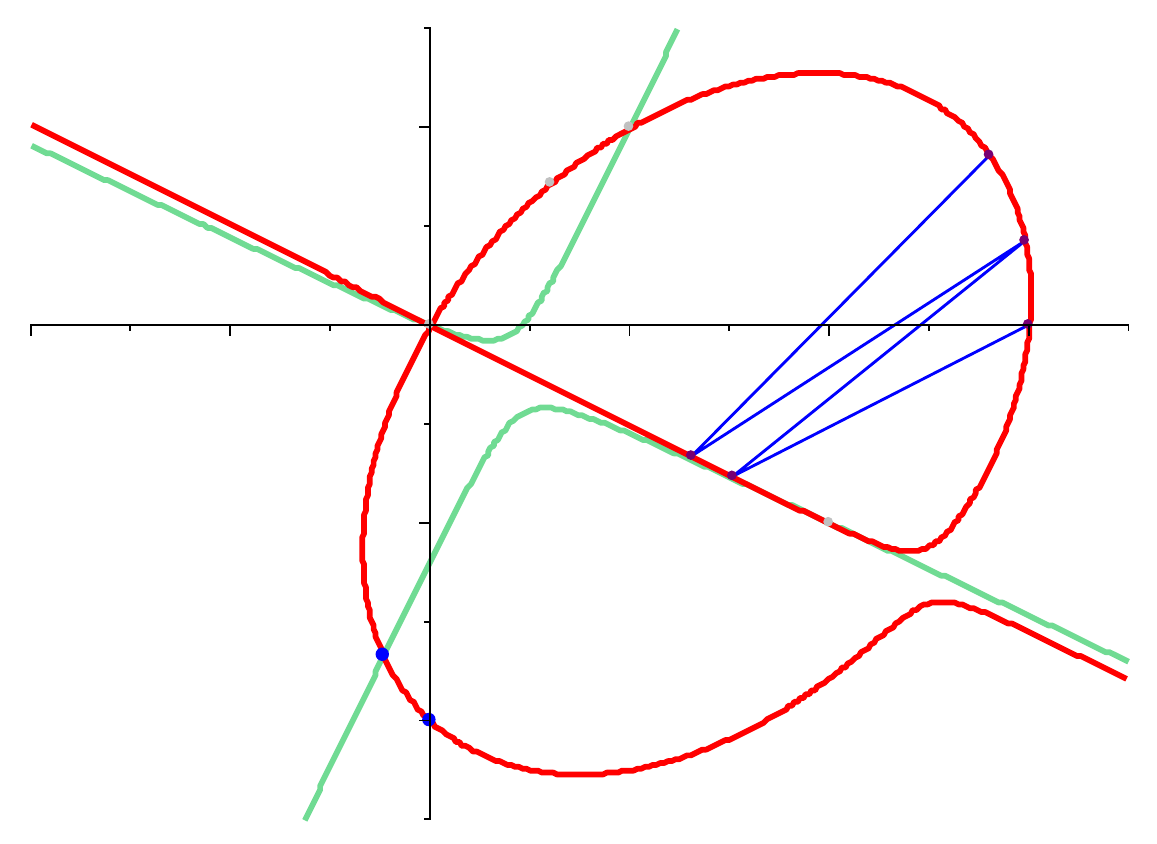}
\put(-190,72){\makebox(0,0)[lb]{\footnotesize$-2$}}
\put(-158,72){\makebox(0,0)[lb]{\footnotesize$-1$}}
\put(-123,72){\makebox(0,0)[lb]{\footnotesize$0$}}
\put(-87,72){\makebox(0,0)[lb]{\footnotesize$1$}}
\put(-55,72){\makebox(0,0)[lb]{\footnotesize$2$}}
\put(-23,72){\makebox(0,0)[lb]{\footnotesize$3$}}
\put(-10,72){\makebox(0,0)[lb]{\footnotesize$u$}}
\put(-123,127){\makebox(0,0)[lb]{\footnotesize$v$}}
\put(-123,111){\makebox(0,0)[lb]{\footnotesize$1$}}
\put(-131,47){\makebox(0,0)[lb]{\footnotesize$-1$}}
\put(-131,16){\makebox(0,0)[lb]{\footnotesize$-2$}}\qquad
\includegraphics[width=6.5cm]{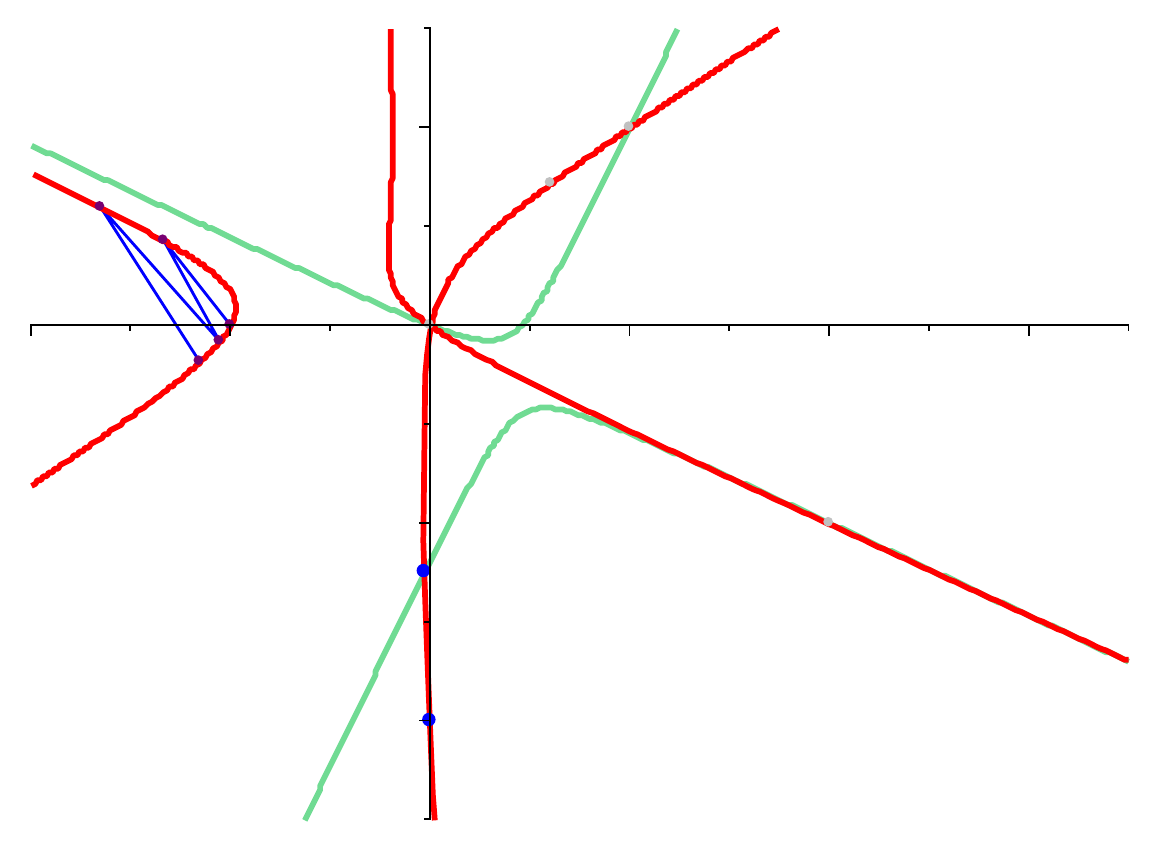}
\put(-190,72){\makebox(0,0)[lb]{\footnotesize$-2$}}
\put(-158,72){\makebox(0,0)[lb]{\footnotesize$-1$}}
\put(-123,72){\makebox(0,0)[lb]{\footnotesize$0$}}
\put(-87,72){\makebox(0,0)[lb]{\footnotesize$1$}}
\put(-55,72){\makebox(0,0)[lb]{\footnotesize$2$}}
\put(-23,72){\makebox(0,0)[lb]{\footnotesize$3$}}
\put(-10,72){\makebox(0,0)[lb]{\footnotesize$u$}}
\put(-123,127){\makebox(0,0)[lb]{\footnotesize$v$}}
\put(-123,111){\makebox(0,0)[lb]{\footnotesize$1$}}
\put(-131,47){\makebox(0,0)[lb]{\footnotesize$-1$}}
\put(-131,16){\makebox(0,0)[lb]{\footnotesize$-2$}}
\caption{The straight blue lines represent the action of $\iota_{b_3}\circ\iota_{Q(-25/2)}$ on the point $(3,0)\in P(-7/10)$ (red curve, left) and on $(-1,0)\in P(13/60)$ (red curve, right). The involution curve $Q(-25/2)$ is shown in aquamarine. The non-base point intersection of $Q(-25/2)$ with $P(-7/10)$ is $(-2195006/9401267, -15699575/9401267)$, and the non-base point intersection of $Q(-25/2)$ with $P(13/60)$ is $(-13617/489844, -1221525/979688)$.}\label{f1}
\end{figure}
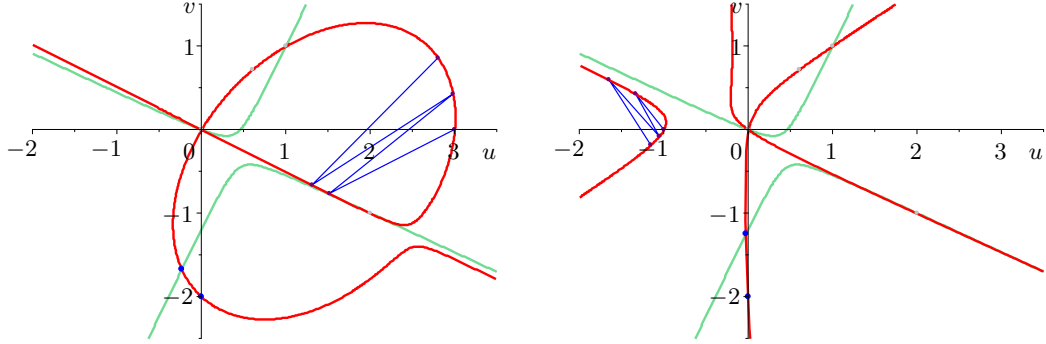

\begin{Example} Consider the pencil $P(C)$ with
\begin{gather}
F=3u^3 + 6u^2v + 9uv^2 + 12v^3 - 23u^2 - 31uv + 24v^2,\nonumber\\
G=(2v + u)\big(35u^2 - 28uv + 26v^2 - 85u + 52v\big).\label{PC}
\end{gather}
It has a double base point at $b_1=(0,0)$, simple base points at $b_1=(2,-1)$, $b_2=(1,1)$, $b_3=(0,-2)$, and $b_4=(331971/549181, 394350/549181)$. All curves have tangent line $T = 20u + 43v + 3=0$ at $b_1$. The pencil
\[
Q(D):=38u^2 + 41uv - 88v^2 + 9u-D(u + 2v)(2u + v - 3)=0
\]
is a pencil of involution curves for $P(C)$, which have simple base points at $b_0$, $b_1$, $b_2$, $b_3$, and tangent line $T$ at $b_1$. An involution is defined for any curve in $Q(D)$, the involution point being
\[
q(C,D)= \big(6 ( 5 D+88 ) ( 13 C-3 )K,-3 ( 65 DC-37 D-
3545 C+901 )K \big)
\]
with $K=\big({D}^{2}+222 DC-55 D+156 C-36\big)\big(
195 C{D}^{3}+43290 {C}^{2}{D}^{2}-23 {D}^{3}-16290 C{D}^{2} + 4572 {C}^{2}D+2256 {D}^{2}+246267 DC+13055022 {C}^{2}-64059 D-7024452 C+939542
\big)^{-1}$.
The expression for $\iota_{Q(D)}(u,v)$ is too large to be included here. It is a rational function, of deg\-ree~5 in~$u$,~$v$, degree $6$ in $D$, and degree $4$ in $C$. By substituting $C=F/G$, one obtains a~rational function of degree~9 in $u$,~$v$, which takes the form
$(P_1P_2/P_4,P_1P_3/P_4)$ with $P_1$, $P_2$, $P_3$, $P_4$ polynomials in~$u$,~$v$ of degree 5, 4, 4, 9 respectively.
We have illustrated the action of~$\iota_{b_3}\circ\iota_{Q(-25/2)}$ on two different curves in $P(C)$ in Figure~\ref{f1}.
\end{Example}

In Figure~\ref{f1} and the subsequent figures, the involution points are indicated by blue dots, and the base points of the pencil by grey dots. The invariant curves are displayed in red, and the involution curves have colors aquamarine or turquoise. The blue straight lines represent the actions of the two involutions, acting on points represented by purple dots.

\begin{figure}[h]\centering
\includegraphics[width=7.5cm]{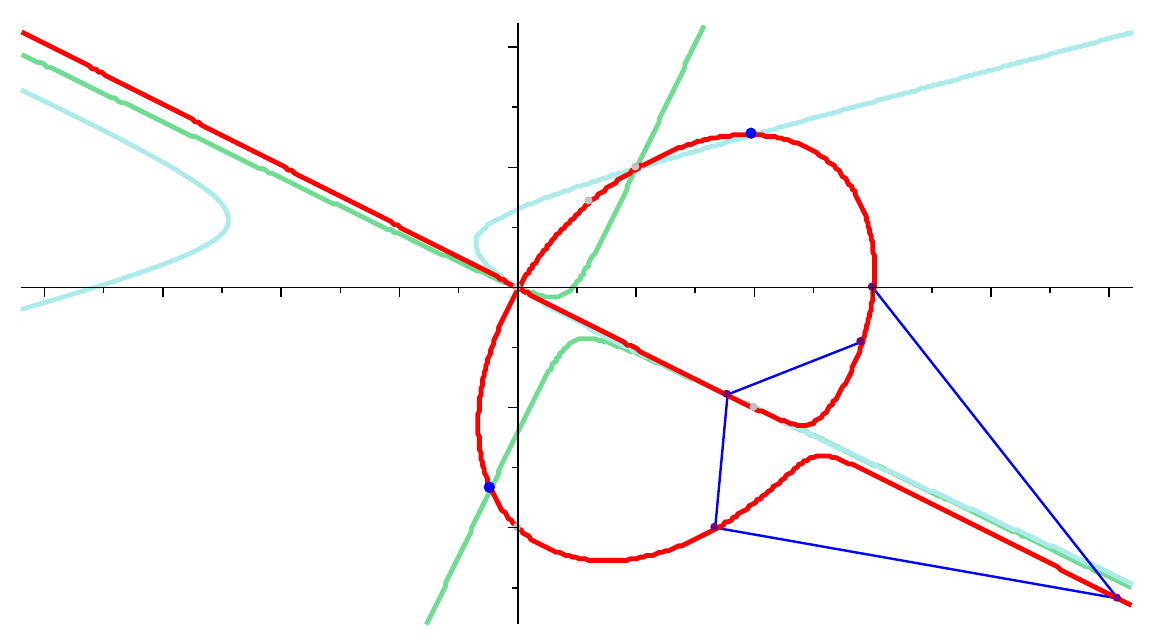}
\put(-216,55){\makebox(0,0)[lb]{\footnotesize$-4$}}
\put(-193,55){\makebox(0,0)[lb]{\footnotesize$-3$}}
\put(-171,55){\makebox(0,0)[lb]{\footnotesize$-2$}}
\put(-149,55){\makebox(0,0)[lb]{\footnotesize$-1$}}
\put(-124,56){\makebox(0,0)[lb]{\footnotesize$0$}}
\put(-99,55){\makebox(0,0)[lb]{\footnotesize$1$}}
\put(-77,55){\makebox(0,0)[lb]{\footnotesize$2$}}
\put(-55,55){\makebox(0,0)[lb]{\footnotesize$3$}}
\put(-34,55){\makebox(0,0)[lb]{\footnotesize$4$}}
\put(-11,55){\makebox(0,0)[lb]{\footnotesize$u$}}
\put(-124,117){\makebox(0,0)[lb]{\footnotesize$v$}}
\put(-125,107){\makebox(0,0)[lb]{\footnotesize$2$}}
\put(-125,85){\makebox(0,0)[lb]{\footnotesize$1$}}
\put(-132,40){\makebox(0,0)[lb]{\footnotesize$-1$}}
\put(-132,18){\makebox(0,0)[lb]{\footnotesize$-2$}}\quad\
\includegraphics[width=7.5cm]{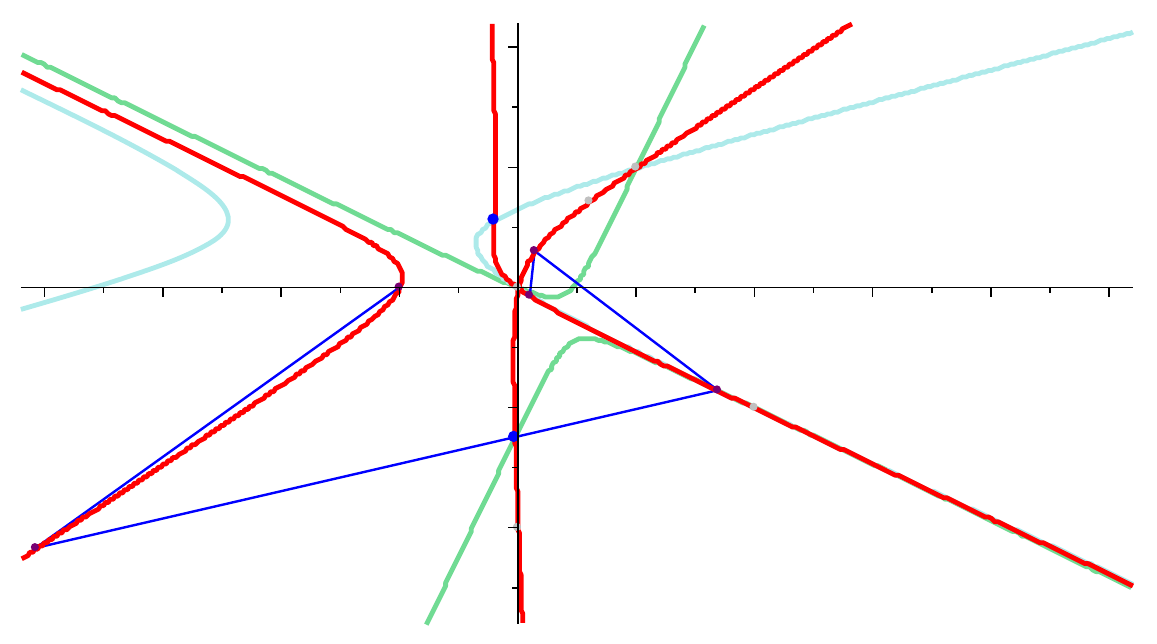}
\put(-216,55){\makebox(0,0)[lb]{\footnotesize$-4$}}
\put(-193,55){\makebox(0,0)[lb]{\footnotesize$-3$}}
\put(-171,55){\makebox(0,0)[lb]{\footnotesize$-2$}}
\put(-149,55){\makebox(0,0)[lb]{\footnotesize$-1$}}
\put(-125,56){\makebox(0,0)[lb]{\footnotesize$0$}}
\put(-99,55){\makebox(0,0)[lb]{\footnotesize$1$}}
\put(-77,55){\makebox(0,0)[lb]{\footnotesize$2$}}
\put(-55,55){\makebox(0,0)[lb]{\footnotesize$3$}}
\put(-34,55){\makebox(0,0)[lb]{\footnotesize$4$}}
\put(-11,55){\makebox(0,0)[lb]{\footnotesize$u$}}
\put(-124,117){\makebox(0,0)[lb]{\footnotesize$v$}}
\put(-125,107){\makebox(0,0)[lb]{\footnotesize$2$}}
\put(-125,85){\makebox(0,0)[lb]{\footnotesize$1$}}
\put(-133,40){\makebox(0,0)[lb]{\footnotesize$-1$}}
\put(-133,18){\makebox(0,0)[lb]{\footnotesize$-2$}}
\caption{The action of $\iota_{Q(-25/2)}\circ\iota_{Q(25/2)}$ on the point $(3,0)\in P(-7/10)$ (left) and on $(-1,0)\in P(13/60)$ (right). The involution curve $Q(-25/2)$ is shown in aquamarine, and the invo\-lution curve $Q(25/2)$ is turquoise. The non-base point intersection of $Q(25/2)$ with $P(-7/10)$ is $(1040561214/525932177, 671760675/525932177)$, and the non-base point intersection of $Q(25/2)$ with $P(13/60)$ is $(-136052/672789, 378550/672789)$.}\label{f3}
\end{figure}
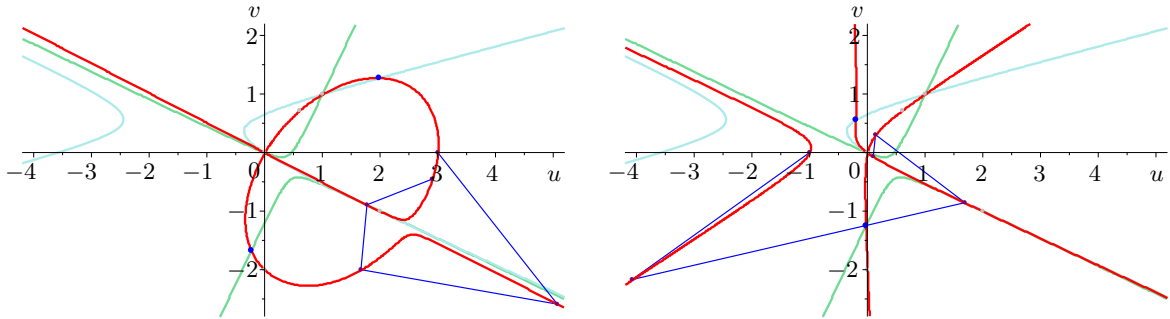

Composing two Manin involutions with distinct involution curves also gives an integrable map of the plane. The action of $\iota_{Q(-25/2)}\circ\iota_{Q(25/2)}$ on the same curves in $P(C)$ is illustrated in~Figure~\ref{f3}.

The previous section, cf.~Table~\ref{BPI}, also showed that the special extra feature is not necessary. For pencils with or without infinitely near base points we have the following possibilities.
\begin{Proposition}\label{P2}
Let $P(C)$ be a pencil of cubic curves, which has $1$ double base point $b_0$ and simple base points $b_1$, $b_2$, $b_3$, $b_4$, $b_5$, which may or may not be infinitely near. Let $Q$ be a curve such that either
\begin{itemize}\itemsep=0pt
\item $\deg(Q)=1$, $b_0$ is a simple point on $Q$, or
\item $\deg(Q)=1$, $2$ simple base point of $P(C)$ are simple points on $Q$, or
\item $\deg(Q)=2$, $5$ simple base points of $P(C)$ are simple points on $Q$, or
\item $\deg(Q)=2$, $4$ base points of $P(C)$ including $b_0$ are simple points on $Q$, or
\item $\deg(Q)=3$, $b_0$ is a double point on $Q$, $4$ other base points of $P(C)$ are simple points on~$Q$.
\end{itemize}
Then $Q$ is an involution curve.
\end{Proposition}

In fact, Proposition~\ref{P1} arises as special cases of the last two items in Proposition~\ref{P2}.

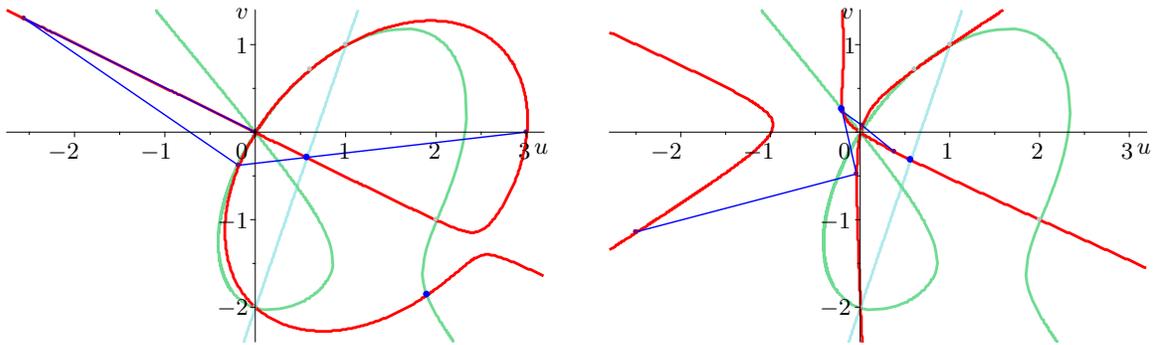
\begin{figure}[h]\centering
\includegraphics[width=7.5cm]{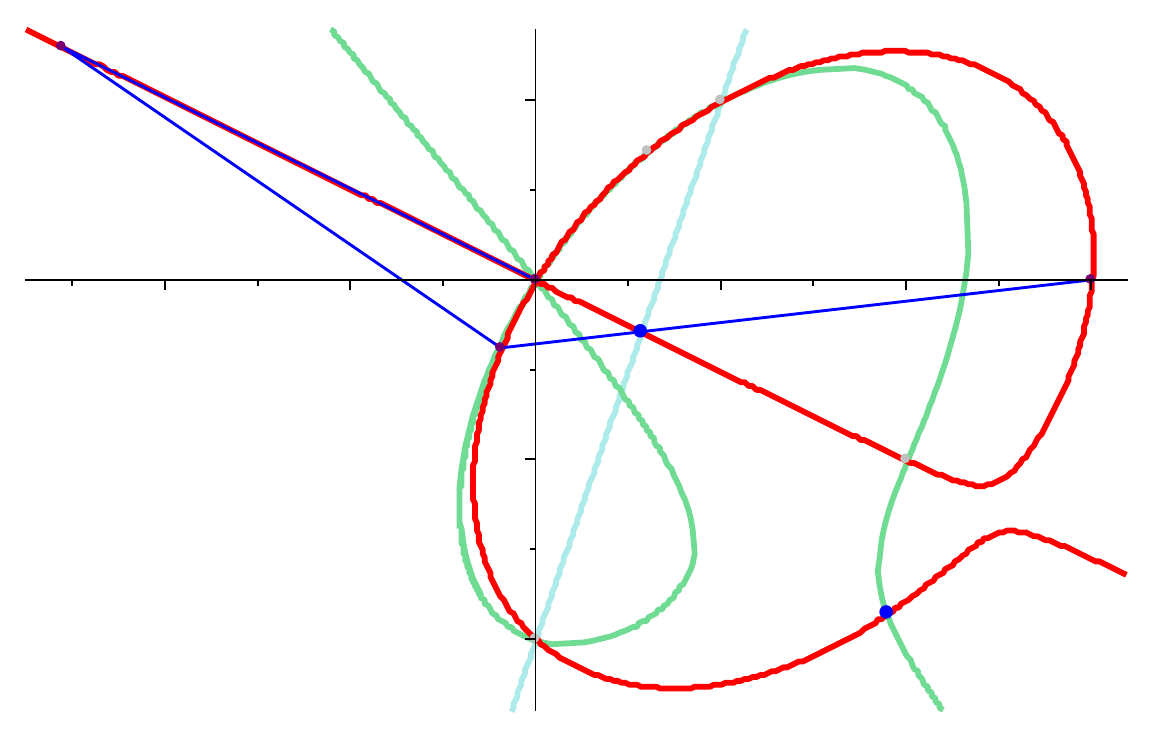}
\put(-193,73){\makebox(0,0)[lb]{\footnotesize$-2$}}
\put(-158,73){\makebox(0,0)[lb]{\footnotesize$-1$}}
\put(-122,74){\makebox(0,0)[lb]{\footnotesize$0$}}
\put(-83,74){\makebox(0,0)[lb]{\footnotesize$1$}}
\put(-49,74){\makebox(0,0)[lb]{\footnotesize$2$}}
\put(-15,74){\makebox(0,0)[lb]{\footnotesize$3$}}
\put(-9,76){\makebox(0,0)[lb]{\footnotesize$u$}}
\put(-122,127){\makebox(0,0)[lb]{\footnotesize$v$}}
\put(-122,114){\makebox(0,0)[lb]{\footnotesize$1$}}
\put(-129,47){\makebox(0,0)[lb]{\footnotesize$-1$}}
\put(-129,14){\makebox(0,0)[lb]{\footnotesize$-2$}}\quad\
\includegraphics[width=7.5cm]{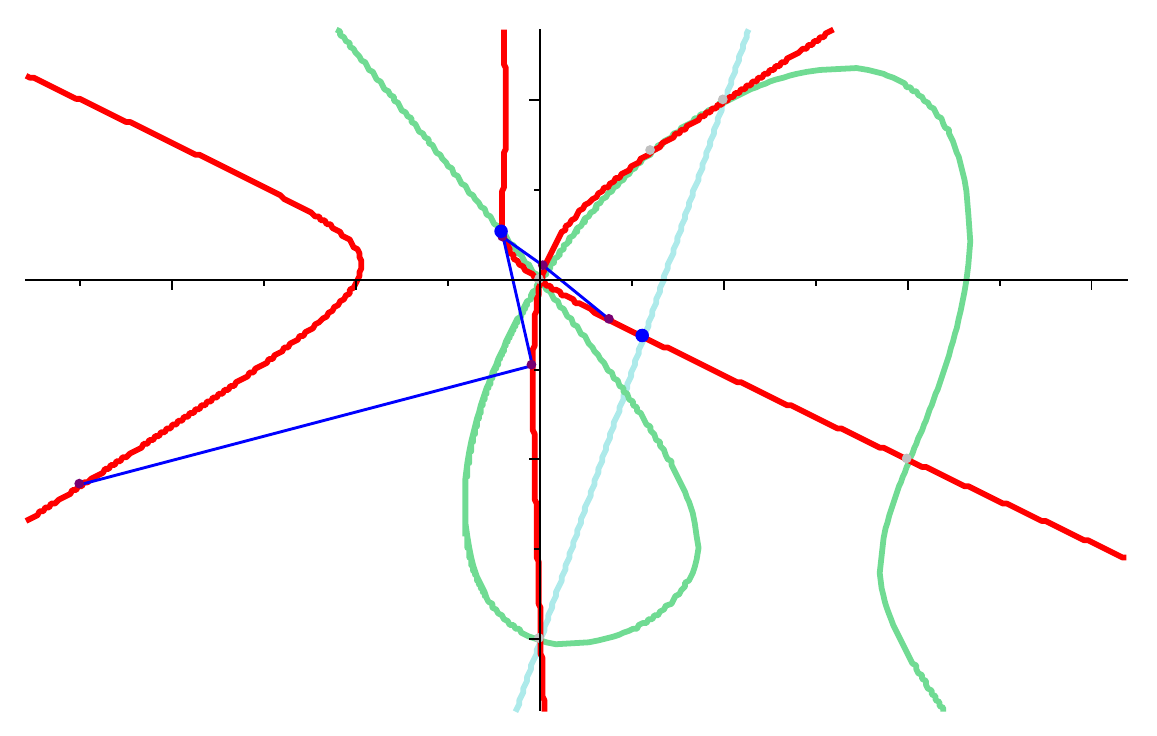}
\put(-193,73){\makebox(0,0)[lb]{\footnotesize$-2$}}
\put(-158,73){\makebox(0,0)[lb]{\footnotesize$-1$}}
\put(-122,74){\makebox(0,0)[lb]{\footnotesize$0$}}
\put(-83,74){\makebox(0,0)[lb]{\footnotesize$1$}}
\put(-49,74){\makebox(0,0)[lb]{\footnotesize$2$}}
\put(-15,74){\makebox(0,0)[lb]{\footnotesize$3$}}
\put(-9,76){\makebox(0,0)[lb]{\footnotesize$u$}}
\put(-121,127){\makebox(0,0)[lb]{\footnotesize$v$}}
\put(-120,114){\makebox(0,0)[lb]{\footnotesize$1$}}
\put(-129,47){\makebox(0,0)[lb]{\footnotesize$-1$}}
\put(-129,14){\makebox(0,0)[lb]{\footnotesize$-2$}}
\caption{The action of $\iota_{Z}\circ\iota_{L}$ on the point $(3,0)\in P(-7/10)$ (left) and on $(-1,0)\in P(13/60)$ (right). The involution curve $Z$ is shown in aquamarine, and the involution curve $L$ is turquoise. The non-base point intersection of $L$ with $P(-7/10)$ is $(304/533, -154/533)$, the non-base point intersection of $Z$ with $P(-7/10)$ is $(1863583907/981888713, -1821490636/981888713)$, the non-base point intersection of $L$ with $P(13/60)$ is $(196/349, -110/349)$, the non-base point intersection of $Z$ with $P(13/60)$ is $(-2223/10798, 5733/21596)$.} \label{f4}
\end{figure}

\begin{Example} The pencil $P(C)$ with (\ref{PC}) admits the involution curves:
\begin{gather*}
L:=3x - y - 2=0,\qquad Z:=5509x^3 + 3032y^3 - 12719x^2 - 1886xy + 6064y^2.
\end{gather*}
The line $L$ intersects $P(C)$ in the base points $b_2$, $b_3$, and the point
$\big(2 \frac {370 C-121}{1295 C-426},-2 \frac {185 C-63}{1295 C-426}\big)$.
The singular cubic $Z$ is one curve in a net of curves which admits $b_0$ as double point and $b_1$, $b_2$, $b_3$, $b_4$ as simple points. The remaining intersection point for $Z$ is
\begin{gather*}
\bigg(4 \frac {3016440 {C}^{3}-12063555 {C}^{2}+13156506 C-2480504}{
6032880 {C}^{3}-26341812 {C}^{2}+31299114 C-2389403},
\\ \qquad
{}-7 \frac{861840 {C}^{3}-5355090 {C}^{2}+8301321 C-1677988}{6032880 {C}^{3}-
26341812 {C}^{2}+31299114 C-2389403}\bigg).
\end{gather*}
The expressions $\iota_{Z}(u,v)$ and $\iota_{L}(u,v)$ are rational functions of degree 5 in $u$,~$v$, and of degree~6, resp.~2 in~$C$. After substitution of $C=F/G$ the expressions are of degree~13 resp.~5 in $u$,~$v$.
The action of the composition of $\iota_{Z}$ and $\iota_{L}$ on selected curves of~$P(C)$ is illustrated in Figure~\ref{f4}.
\end{Example}

Other classes of involution curves, of degree higher than 3 exist, see Appendix~\ref{appendixB}.

\subsection{Involution curves for non-singular cubic pencils}
It is possible to construct involution curves for non-singular cubic pencils.

{\sloppy\begin{Proposition}
Let $P(C)$ be a non-singular pencil of cubic curves, with base points $b_0,b_1,\ldots,b_8$. Let $Q$ be a curve of degree $3$ such that $b_0$ is a double point on $Q$ and six base points of $P(C)$ are simple points on~$Q$. Then $Q$ is an involution curve.
\end{Proposition}

}

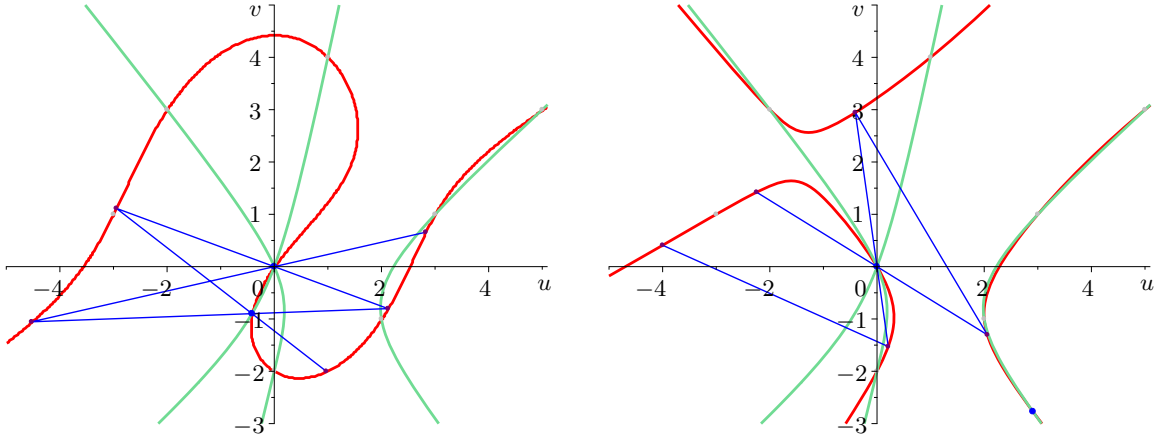
\begin{figure}[h]\centering
\includegraphics[width=7.5cm]{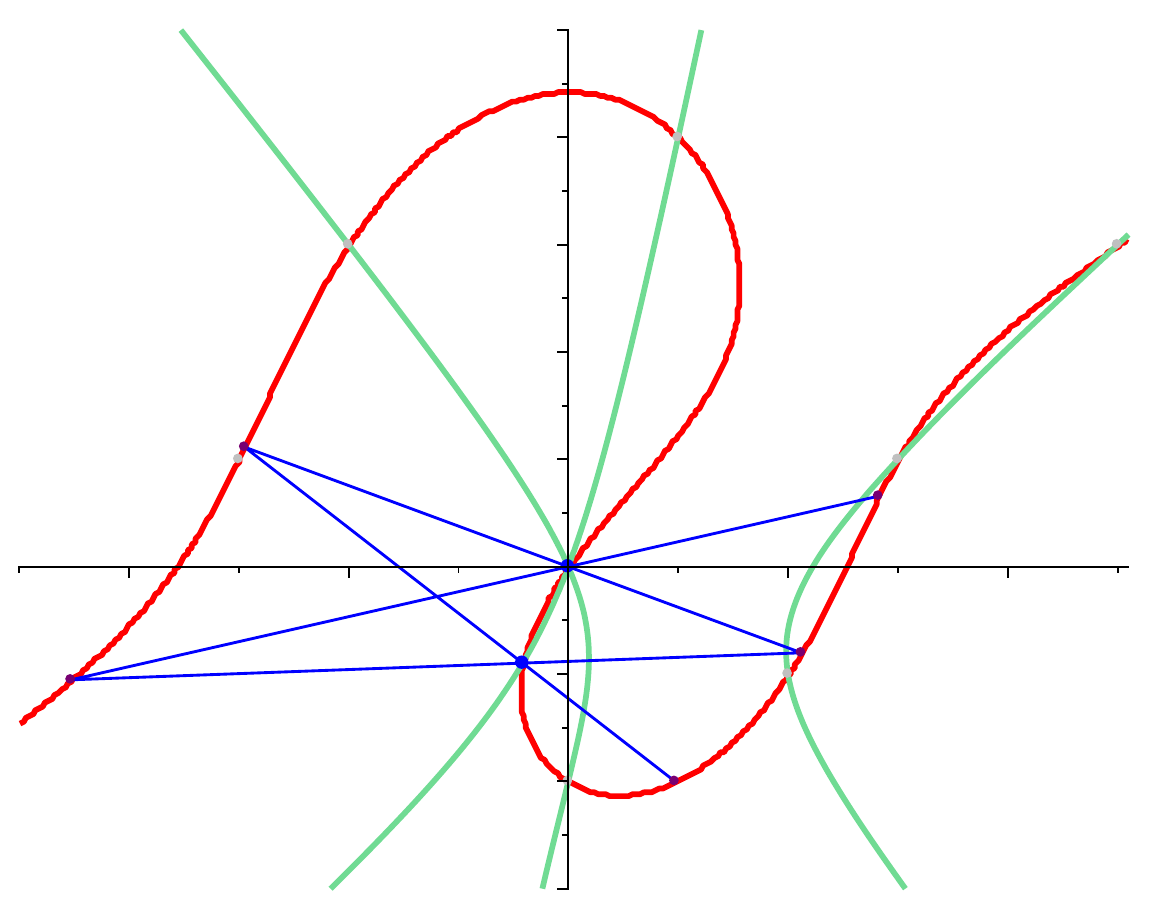}
\put(-200,53){\makebox(0,0)[lb]{\footnotesize$-4$}}
\put(-159,53){\makebox(0,0)[lb]{\footnotesize$-2$}}
\put(-117,54){\makebox(0,0)[lb]{\footnotesize$0$}}
\put(-71,54){\makebox(0,0)[lb]{\footnotesize$2$}}
\put(-31,54){\makebox(0,0)[lb]{\footnotesize$4$}}
\put(-9,56){\makebox(0,0)[lb]{\footnotesize$u$}}
\put(-118,160){\makebox(0,0)[lb]{\footnotesize$v$}}
\put(-118,141){\makebox(0,0)[lb]{\footnotesize$4$}}
\put(-118,121){\makebox(0,0)[lb]{\footnotesize$3$}}
\put(-118,101){\makebox(0,0)[lb]{\footnotesize$2$}}
\put(-118,82){\makebox(0,0)[lb]{\footnotesize$1$}}
\put(-124,41){\makebox(0,0)[lb]{\footnotesize$-1$}}
\put(-124,21){\makebox(0,0)[lb]{\footnotesize$-2$}}
\put(-124,1){\makebox(0,0)[lb]{\footnotesize$-3$}}\quad\
\includegraphics[width=7.5cm]{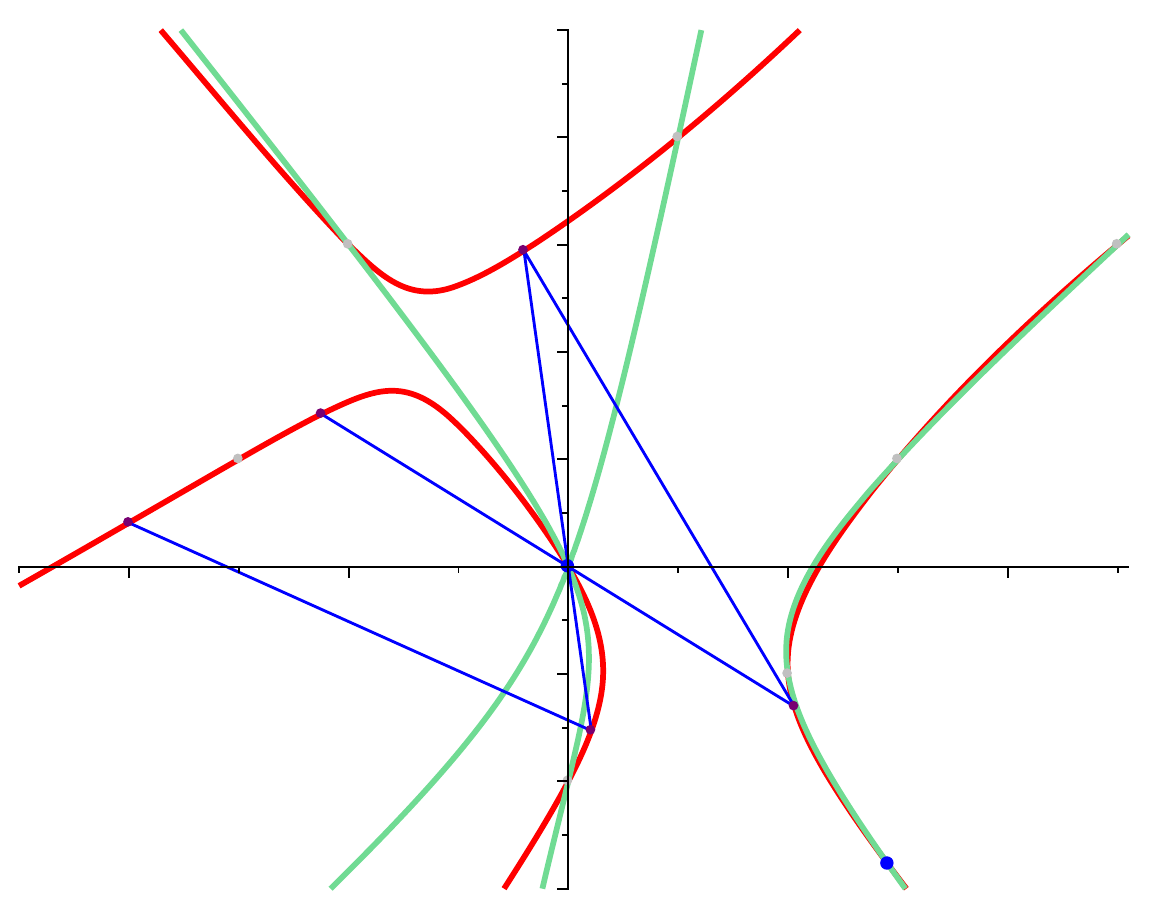}
\put(-200,53){\makebox(0,0)[lb]{\footnotesize$-4$}}
\put(-159,53){\makebox(0,0)[lb]{\footnotesize$-2$}}
\put(-117,54){\makebox(0,0)[lb]{\footnotesize$0$}}
\put(-71,54){\makebox(0,0)[lb]{\footnotesize$2$}}
\put(-31,54){\makebox(0,0)[lb]{\footnotesize$4$}}
\put(-9,56){\makebox(0,0)[lb]{\footnotesize$u$}}
\put(-118,160){\makebox(0,0)[lb]{\footnotesize$v$}}
\put(-118,141){\makebox(0,0)[lb]{\footnotesize$4$}}
\put(-118,121){\makebox(0,0)[lb]{\footnotesize$3$}}
\put(-118,101){\makebox(0,0)[lb]{\footnotesize$2$}}
\put(-118,82){\makebox(0,0)[lb]{\footnotesize$1$}}
\put(-124,41){\makebox(0,0)[lb]{\footnotesize$-1$}}
\put(-124,21){\makebox(0,0)[lb]{\footnotesize$-2$}}
\put(-124,1){\makebox(0,0)[lb]{\footnotesize$-3$}}
\caption{The action of $\iota_{b_0}\circ\iota_{Q}$ on the point $(97/100,-2)\in P(-37210300/12948573)$ (left) and on $(-4,409/1000)\in P(-17169083242044/13796901723833)$ (right), where the pencil $P(C)$ given by (\ref{NSPC}). The involution curve $Q$, given by (\ref{icQ}), is shown in aquamarine. Its non-base point intersections with the two curves of $P(C)$ are given in Appendix~\ref{appendixC}.} \label{f5}
\end{figure}

\begin{Example}
We construct a pencil of cubic curves with finite base points
\begin{gather}
b_0=(0,0),\qquad
b_1=(2,-1),\qquad
b_2=(1,4),\qquad
b_3=(0,-2),\nonumber
\\
b_4=(5,3),\qquad
b_5=(3,1),\qquad
b_6=(-2,3),\qquad
b_7=(-3,1).\label{BP8}
\end{gather}
This fixes a pencil $P(C)$ with
\begin{gather}
F=135u^2v - 1276uv^2 + 1436v^3 + 1173u^2 + 2737uv - 2488v^2 - 1461u - 10720v,\nonumber
\\
G=391u^3 - 772u^2v + 673uv^2 - 777v^3 - 1173uv + 1539v^2 - 3019u + 6186v, \label{NSPC}
\end{gather}
which has 9-th base point
\begin{gather}\label{BP9}
b_8=\bigg(\frac{126933249}{5530213}, \frac{75665173}{5530213}\bigg).
\end{gather}
The following involution curve is the unique curve which has a double point at $b_0$ and simple points $b_1$, $b_2$, $b_3$, $b_4$, $b_5$, $b_6$ in common with $P(C)$:
\begin{equation} \label{icQ}
Q:=597u^3 - 310u^2v - 481uv^2 + 104v^3 - 1329u^2 - 79uv + 208v^2=0.
\end{equation}
The unique non-base point intersection between $Q$ and $P(C)$ is given by
\begin{gather*}
\left({-}598 {\frac {\begin{matrix}5174263414217 {C}^{3}+26456591132843 {C}^{2}\\+
44024545626872 C+23947542820608\end{matrix}}{\begin{matrix}2293729901271491 {C}^{3}+
7965008023759238 {C}^{2}\\+8157021862775051 C+2203322144658228\end{matrix}}},\right.
\\ \hspace*{6mm}
\left.-3 {\frac {\begin{matrix}1636188291089981 {C}^{3}+7577888329154480 {C}^{2}\\+
11109497069357891 C+5212332367047960\end{matrix}}{\begin{matrix}2293729901271491 {C}^{3}+
7965008023759238 {C}^{2}\\+8157021862775051 C+2203322144658228\end{matrix}}} \right)\!.
\end{gather*}
The rational function $\iota_{Q}(u,v)$ is of degree 5 in $u$,~$v$, and of degree 6 in $C$. After substitution of~$C=F/G$ the expression has degree 14 in $u$,~$v$.
The action of $\iota_{b_0}\circ\iota_Q$ on two curves of the pencil is illustrated in Figure~\ref{f5}.
\end{Example}

\begin{Proposition}[Bertini involution]
Let $P(C)$ be a non-singular pencil of cubic curves, with base points $b_0,\ldots,b_8$. Let $Q$ be a curve of degree $4$ such that $b_0$ is a triple point on $Q$ and $b_1,\ldots,b_8$ are simple points on~$Q$. Then $Q$ is an involution curve.
\end{Proposition}

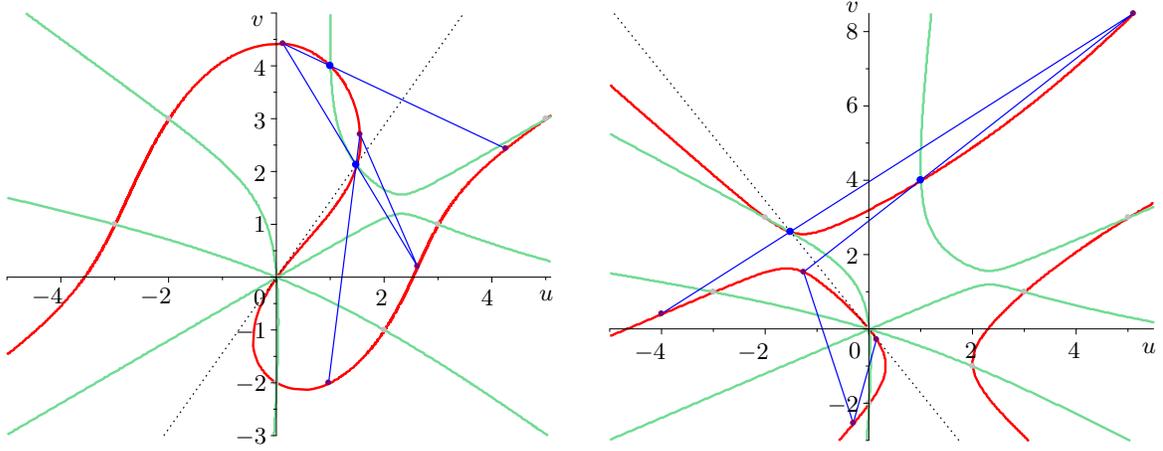
\begin{figure}[h]\centering
\includegraphics[width=7.5cm]{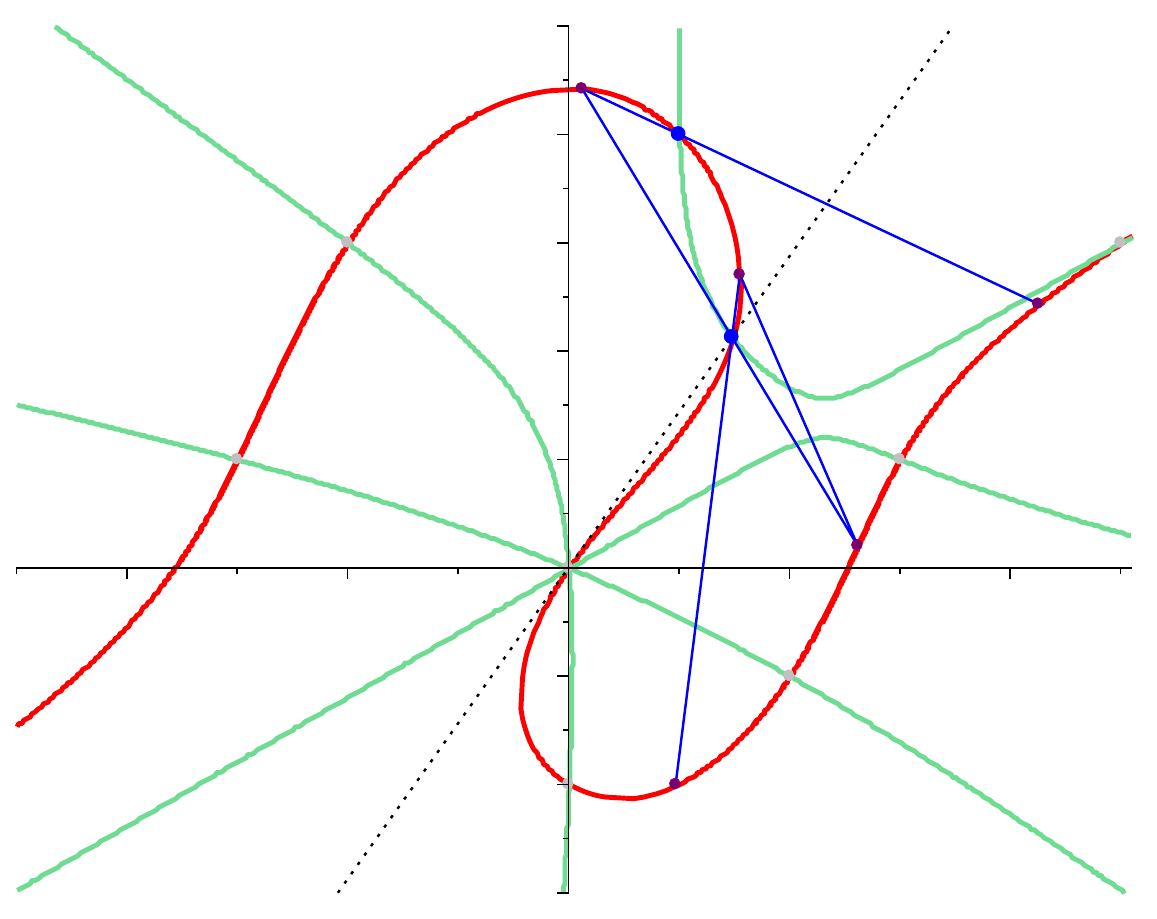}
\put(-201,54){\makebox(0,0)[lb]{\footnotesize$-4$}}
\put(-160,54){\makebox(0,0)[lb]{\footnotesize$-2$}}
\put(-117,54){\makebox(0,0)[lb]{\footnotesize$0$}}
\put(-71,54){\makebox(0,0)[lb]{\footnotesize$2$}}
\put(-31,54){\makebox(0,0)[lb]{\footnotesize$4$}}
\put(-9,56){\makebox(0,0)[lb]{\footnotesize$u$}}
\put(-118,160){\makebox(0,0)[lb]{\footnotesize$v$}}
\put(-117,141){\makebox(0,0)[lb]{\footnotesize$4$}}
\put(-117,121){\makebox(0,0)[lb]{\footnotesize$3$}}
\put(-117,101){\makebox(0,0)[lb]{\footnotesize$2$}}
\put(-117,82){\makebox(0,0)[lb]{\footnotesize$1$}}
\put(-124,41){\makebox(0,0)[lb]{\footnotesize$-1$}}
\put(-124,21){\makebox(0,0)[lb]{\footnotesize$-2$}}
\put(-124,1){\makebox(0,0)[lb]{\footnotesize$-3$}}\quad\
\includegraphics[width=7.5cm]{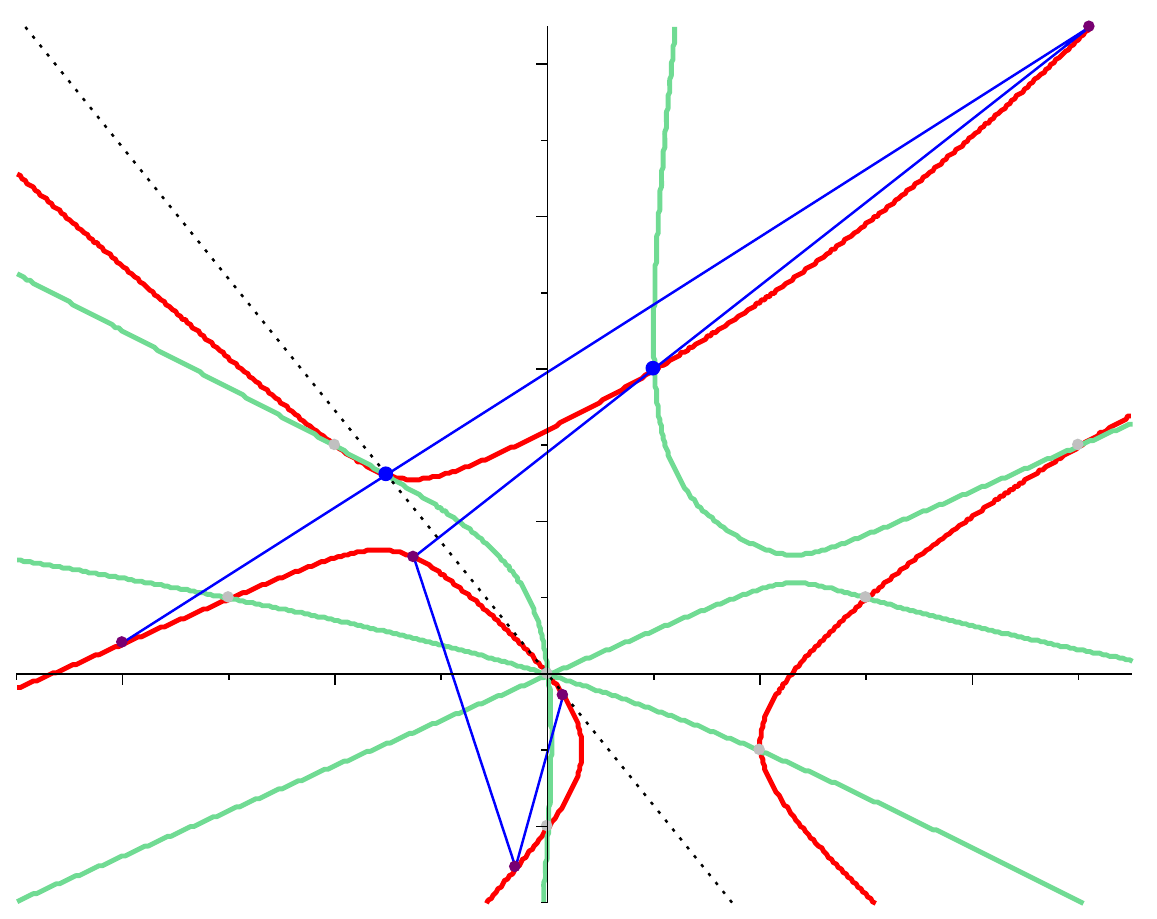}
\put(-201,33){\makebox(0,0)[lb]{\footnotesize$-4$}}
\put(-162,33){\makebox(0,0)[lb]{\footnotesize$-2$}}
\put(-120,34){\makebox(0,0)[lb]{\footnotesize$0$}}
\put(-76,34){\makebox(0,0)[lb]{\footnotesize$2$}}
\put(-37,34){\makebox(0,0)[lb]{\footnotesize$4$}}
\put(-9,36){\makebox(0,0)[lb]{\footnotesize$u$}}
\put(-121,165){\makebox(0,0)[lb]{\footnotesize$v$}}
\put(-121,155){\makebox(0,0)[lb]{\footnotesize$8$}}
\put(-121,127){\makebox(0,0)[lb]{\footnotesize$6$}}
\put(-121,99){\makebox(0,0)[lb]{\footnotesize$4$}}
\put(-121,70){\makebox(0,0)[lb]{\footnotesize$2$}}
\put(-128,13){\makebox(0,0)[lb]{\footnotesize$-2$}}
\caption{The action of $\iota_{b_3}\circ\iota_{B}$ on the same points as in figure \ref{f5}. The involution curve $B$, given by (\ref{icB}), is shown in aquamarine. Its non-base point intersections with the two curves of $P(C)$ are given in Appendix~\ref{appendixC}. The tangents at $b_0$, which are given by $335692483u - 233693640v=0$ (left) and $184119528711639u + 106636413156536v=0$ (right), are drawn with dotted black lines.}\label{f6}
\end{figure}

\begin{Example}
Consider the pencil $P(C)$ with (\ref{NSPC}). The unique curve of degree 4 which has a~triple point at $b_0$ and simple points at $b_1,\ldots,b_8$ given by (\ref{BP8}) and (\ref{BP9}),
\begin{gather*}
B:=1461 {x}^{4}+6793 {x}^{3}y-6663 {x}^{2}{y}^{2}-15727 {y}^{3}x+1416
 {y}^{4}-9057 {x}^{3}-6958 y{x}^{2}\nonumber
\\ \hphantom{B:=}
{}+36103 {y}^{2}x+2832 {y}^{3}=0,\label{icB}
\end{gather*}
\looseness=1
is an involution curve. The unique non-base point intersection between $B$ and $P(C)$ is gi\-ven~by
\begin{gather}
\left(\frac{2}{17} {\frac {\begin{matrix}2215543962789 C^4 + 35968027317846 C^3 + 117454744627949 C^2 \\+ 131413745083336 C + 42208499157120\end{matrix}}{\begin{matrix}
106600301109 C^4 + 1562236994153 C^3 + 3806116035387 C^2\\ + 2550071307523 C + 301292311428\end{matrix}}}
,\right. \nonumber
\\[1ex] \hspace*{3mm}
\left.\frac{1}{17} \frac {\begin{matrix}2162537091387 C^4 + 30313406379925 C^3 + 45123416722925 C^2 \\- 5407306540485 C - 11504965908312\end{matrix}}{\begin{matrix}106600301109 C^4 + 1562236994153 C^3 + 3806116035387 C^2\\ + 2550071307523 C + 301292311428\end{matrix}}
\right)\!.\label{unbpi}
\end{gather}

\looseness=1
The rational function $\iota_{B}(u,v)$ is of degree 5 in $u$,~$v$, and of degree 8 in $C$. After substitution of $C=F/G$ the expression has degree 16 in $u$,~$v$. The action of $\iota_{b_3}\circ\iota_B$ on two curves of the pencil is illustrated in Figure~\ref{f6}. We have also shown the fact that the tangent to the cubic,
\begin{gather*}
(3019 C-1461) \big(716309073 {C}^{3}+10387523662 {C}^{2}+19973364953 C+7874719992\big) x
\\ \qquad
{}-2 (3093 C+5360) \big(716309073 {C}^{3}+10387523662 {C}^{2}+19973364953 C
\\ \qquad \hphantom{-2 \big(}
{}+7874719992\big) y=0
\end{gather*}
intersects the quartic in (\ref{unbpi}), cf.~\cite{Moo}.
\end{Example}

For non-singular cubic pencils, many other classes of involution curves exist, cf.\ Appendix~\ref{appendixB}.

\section{Measure preservation}
We verified for each involution considered explicitly in the previous section, that it is anti measure-preserving with density $1/F$. We expect this is true for any involution $\iota_Q$ with involution curve $Q$, which preserves a pencil of cubic curves $F-CG=0$.

\appendix

\section[Involution points for the map gamma]
{Involution points for the map $\boldsymbol\gamma$}
\label{appendixA}
The formulas for the intersections of the pencil with the involution curves for the map obtained by open reduction of the Q1$(\delta=0)$ equation, i.e., the parametrisations of the involution points, are given by
\begin{gather*}
h_1=\left({\frac {\big(Ca-Cb+{a}^{2}-{b}^{2}\big)a}{(-C{a}^{2}+Ca
b-2 {a}^{3}+2 b{a}^{2}+{C}^{2}+4 Ca+4 {a}^{2}) b}},
-\frac{a(C+a+b)}{b\big({-}{a}^{2}+ab+C+2 a\big)}\right),
\\
h_2=\left({-}\frac {{C}^{2}+Ca+3 Cb+2 ab+2 {b}^{2}}{C{a}^{2}-Cab+2 b{a}^{2}-
2 a{b}^{2}+{a}^{2}-2 ab+{b}^{2}},-\frac {C+a+b}{Ca+2 ab+a-b}\right),
\\
h_3=\left(\frac{A a}{D},-\frac { \big( {C}
^{2}+2 Ca+2 Cb+{a}^{2}+2 ab+{b}^{2} \big) a}{C{a}^{3}-C{a}^{2}b+3
 {a}^{3}b-4 {a}^{2}{b}^{2}+a{b}^{3}-Cab+C{b}^{2}+{a}^{3}-4 b{a}^{2}
+3 a{b}^{2}}\right),
\\
h_4=\left({-}\frac {Cab-C{b}^{2}+b{a}^{2}-{b}^{3}+{C}^{2}+Ca+3 Cb+2 ab+2 {b}^
{2}}{B},{\frac {C+a+b}{ab-{b}^{2}-a+b}}\right),
\\
k_1=\left({-}\frac {({C}^{2}a-{C}^{2}b+2 C{a}^{2}-2 C{b}^{2}+{a}^{3}+b{a}^{2}-a{b}^{2}-{b}^{3}) {b}^{2}}{a E},\right.
\\ \left.\hphantom{k_1=}
{}-\frac {\big({C}^{2}+2 Ca+2 Cb+{a}^{2}+2 ab+{b}^{2}\big)b}{C{a}^{2}b-Ca{b}^{2}+3 {a}^{2}{b}^{2} -4 a{b}^{3}+{b}^{4}+{C}^{2}a+5 Cab-C{b}^{2}+6 a{b}^{2}-2 {b}^{3}}\right),
\\
k_2=\left(\frac {\big({C}^{3}+4 {C}^{2}a+2 {C}^{2}b+5 C{a}^{2}+6 Cab+C{b}^{2}+2 {a}^{3} +4 b{a}^{2}+2 a{b}^{2}\big) b}{F},\right.
\\ \left.\hphantom{k_2=}
{}-\frac{\big({C}^{2}+2 Ca+2 Cb+{a}^{2}+2 ab+{b}^{2} \big) a}{{C}^{2}{b}^{2}-C{a}^{2}b+5 Ca{b}^{2}-2 {a}^{3}b +6 {a}^{2}{b}^{2}-Cab+C{b}^{2}+{a}^{3}-4 b{a}^{2}+3 a{b}^{2}}\right),
\\
k_3=\left({-}\frac{b(C+a+b)}{a(ab-{b}^{2}-a+b)},{\frac {C+a+b}{ab-{b}^{2}-a+b}}\right),
\\
k_4=\left(\frac{b G}{H},-\frac {\big({C}^{2}+2 Ca+2 Cb+{a}^{2}+2 ab+{b}^{2}\big)a} {C{a}^{3}-C{a}^{2}b+3 {a}^{3}b-4 {a}^{2}{b}^{2}+a{b}^{3}-Cab+C{b}^{2}+{a}^{3} -4 b{a}^{2}+3 a{b}^{2}}\right),
\end{gather*}
with
\begin{gather*}
A={C}^{3}a+2 {C}^{2}{a}^{2}+5 {C}^{2}ab-{C}^{2}{b}^{2
}+C{a}^{3}+8 C{a}^{2}b+5 Ca{b}^{2}-2 C{b}^{3}+3 {a}^{3}b\\
\hphantom{A=}{}+5 {a}^{2
}{b}^{2}+a{b}^{3}-{b}^{4},\\
B=Ca{b}^{2}-C{b}^{3}+2 {a}^{2}{b}^{2}-2 a{b}^{3}-Cab+C{b}^{2}-3
b{a}^{2}+4 a{b}^{2}-{b}^{3}+{a}^{2}-2 ab+{b}^{2},\\
D={C}^{2}{a}^{3}b-{C}^{2}{a}^{2}{b}
^{2}-C{a}^{5}+4 C{a}^{4}b-4 C{a}^{2}{b}^{3}+Ca{b}^{4}-3 {a}^{5}b+13
 {a}^{4}{b}^{2}-13 {a}^{3}{b}^{3}\\
 \hphantom{D=}{}
 +3 {a}^{2}{b}^{4}-{C}^{2}a{b}^{2}+
{C}^{2}{b}^{3}+2 C{a}^{3}b-8 C{a}^{2}{b}^{2}+6 Ca{b}^{3}-{a}^{5}+7
 {a}^{4}b-15 {a}^{3}{b}^{2}+9 {a}^{2}{b}^{3},\\
E={C}^{2}{a}^{2}b-{C}^{
2}a{b}^{2}+5 C{a}^{2}{b}^{2}-6 Ca{b}^{3}+C{b}^{4}+6 {a}^{2}{b}^{3}-
8 a{b}^{4}+2 {b}^{5}+{C}^{3}a+7 {C}^{2}ab\\
 \hphantom{E=}{}
-{C}^{2}{b}^{2}+16 Ca{b}^
{2}-4 C{b}^{3}+12 a{b}^{3}-4 {b}^{4},\\
F={C}^{2}a{b}^{2}-{
C}^{2}{b}^{3}-C{a}^{3}b+6 C{a}^{2}{b}^{2}-5 Ca{b}^{3}-2 {a}^{4}b+8
 {a}^{3}{b}^{2}-6 {a}^{2}{b}^{3}-C{a}^{2}b+2 Ca{b}^{2}\\
 \hphantom{F=}{}
 -C{b}^{3}+{a}
^{4}-5 {a}^{3}b
+7 {a}^{2}{b}^{2}-3 a{b}^{3},\\
G=-{C}^{2}{a}^{2}+{C}^{2}ab-2 C{a}^{3}+2 Ca{b}^{2}-{
a}^{4}-{a}^{3}b+{a}^{2}{b}^{2}+a{b}^{3}+{C}^{3}+4 {C}^{2}a+2 {C}^{2}
b\\
 \hphantom{G=}{}
+5 C{a}^{2}+6 Cab+C{b}^{2}+2 {a}^{3}+4 b{a}^{2}+2 a{b}^{2},\\
H={C}^{2}{a}^{4}-{C}^{2}{a}^{3}b+5 C{a}^{4}b-6 C{a}^{3}{b}^
{2}+C{a}^{2}{b}^{3}+6 {a}^{4}{b}^{2}-8 {a}^{3}{b}^{3}+2 {a}^{2}{b}^
{4}-{C}^{2}{a}^{2}b\\
 \hphantom{H=}{}
+{C}^{2}a{b}^{2}+2 C{a}^{4}-6 C{a}^{3}b+2 C{a}^{
2}{b}^{2}+2 Ca{b}^{3}+5 {a}^{4}b-15 {a}^{3}{b}^{2}+11 {a}^{2}{b}^{
3}-a{b}^{4}\\
 \hphantom{H=}{}
-C{a}^{2}b+2 Ca{b}^{2}-C{b}^{3}+{a}^{4}-5 {a}^{3}b+7 {a}
^{2}{b}^{2}-3 a{b}^{3}.
\end{gather*}

\section{Higher degree involution curves}\label{appendixB}
An irreducible curve $Q$, of degree $M$, with $m_i$ singular points of multiplicity $i\geq 2$ has non-negative genus, i.e.,
\begin{gather} \label{gge0}
\frac{(M-1)(M-2)}2-\sum_i m_i\frac{i(i-1)}2 \geq 0.
\end{gather}
One can recursively generate, for each $M$, all sequences $m=(m_2,m_3,\ldots,m_r)$ such that (\ref{gge0}) is satisfied.
For given $m$, define the sequence
\[
s(m)=(s_1,s_2,\ldots)=\Big(\overbrace{r,\ldots,r}^{\text{$m_r$ times}},\ldots,\overbrace{2,\ldots,2}^{\text{$m_2$ times}},1,1,\ldots\Big).
\]
One can easily verify for each $m$ whether
\begin{gather}
2s_1+\sum_{i=2}^6 s_i \geq 3M-1\label{A},
\end{gather}
or
\begin{gather}
\sum_{i=1}^9 s_i \geq 3M-1.\label{B}
\end{gather}
Condition (\ref{A}) is a necessary condition for $Q$ to be an involution curve for a cubic pencil with a singular base point, whereas (\ref{B}) is a necessary condition for $Q$ to be an involution curve for a non-singular cubic pencil.

\subsection*{Higher degree involution curves for singular pencils}
Using Maple \cite{MAP}, we have observed (for $M\leq 14$) that $2s_1+\sum_{i=2}^6 s_i \leq 3M-1$ for all $m$, and that equality holds when, with $M>3$,
\begin{itemize}\itemsep=0pt
\item $m=(3)$ or $m=(0,1)$ for $M=4$,
\item $m=(6)$ or $m=(3,1)$ for $M=5$,
\item $m=(4,2)$ or $m=(4,0,1)$ for $M=6$,
\item $m=\big(\overbrace{0,\ldots,0}^{\text{$k$ times}},3,2,\overbrace{0,\ldots,0}^{\text{$k$ times}},1\big)$ or $m=\big(\overbrace{0,\ldots,0}^{\text{$k$ times}},5,0,0,\overbrace{0,\ldots,0}^{\text{$k$ times}},1\big)$ for $M=7+3k$,
\item $m=\big(\overbrace{0,\ldots,0}^{\text{$k$ times}},0,5,\overbrace{0,\ldots,0}^{\text{$k$ times}},1\big)$ or $m=\big(\overbrace{0,\ldots,0}^{\text{$k$ times}},2,3,0,\overbrace{0,\ldots,0}^{\text{$k$ times}},1\big)$ for $M=8+3k$,
\item $m=\big(\overbrace{0,\ldots,0}^{\text{$k$ times}},0,4,1,\overbrace{0,\ldots,0}^{\text{$k$ times}},1\big)$ or $m=\big(\overbrace{0,\ldots,0}^{\text{$k$ times}},1,4,0,0,\overbrace{0,\ldots,0}^{\text{$k$ times}},1\big)$ for $M=9+3k$.
\end{itemize}
It is easily verified that equality holds for given $m$ for all $M\geq 7$, e.g., for $M=7+3k$ we have
\[
3(k+2)+2(k+3)+2(2k+4)=3(7+3k)-1.
\]

\subsection*{Higher degree involution curves for non-singular pencils}
Using Maple \cite{MAP}, we have observed (for $M\leq 14$) that $\sum_{i=1}^9 s_i \leq 3M$ for all $m$. For $M>3$, when $3$ divides $M$ there are 3 cases where equality holds and when $3$ does not divide $M$ there is one such $m$. When $M<9$ these cases give rise to an involution curve, when $M>9$ they don't. When $M=9$ we have
$m=(1,7,1)$, which gives rise to an involution curve $Q$ which has $7$ triple points, 1 quadruple point and 1 simple point at base points of $P(C)$, as $1\cdot1+7\cdot3+1\cdot4=9\cdot3-1$. When $M=9$, we also have $m=(0,9)$ and $m=(1,9)$, which do not give rise to an involution curve. All cases for $3<M<9$ are
\begin{gather*}
M=4,\quad m=(3); \qquad\quad M=5,\quad m=(6); \qquad M=6,\quad m=(9),(10),(7,1); \\
M=7,\quad m=(6,3); \qquad M=8,\quad m=(3,6).
\end{gather*}
The number of sub-maximal cases, where $\sum_{i=1}^9 s_i = 3M-1$, which all correspond to a class of involution curves, seems to be growing linearly with $M$. All cases where $3< M\leq 9$ are
\begin{gather*}
M=4,\quad m=(2), (0, 1);\\
M=5,\quad m= (5), (3, 1);\\
M=6,\quad m= (8), (6, 1), (4, 2);\\
M=7,\quad m= (7, 2), (8, 2), (9, 2), (5, 3), (3, 4), (8, 0, 1), (9, 0, 1), (6, 1, 1);\\
M=8,\quad m= (4, 5), (5, 5), (6, 5), (2, 6), (0, 7), (5, 3, 1), (6, 3, 1), (3, 4, 1), (6, 1, 2);\\
M=9,\quad m= (1, 8), (2, 8), (3, 8), (4, 8), (2, 6, 1), (3, 6, 1), (4, 6, 1), (0, 7, 1), (3, 4, 2), (4, 4, 2),
\\ \hphantom{M=9,\quad m=}
 (1, 5, 2), (4, 2, 3), (3, 5, 0, 1).
\end{gather*}

\section{The non-base point intersections in Figures~\ref{f5} and \ref{f6}}\label{appendixC}
In Figures~\ref{f5}, the non-base point intersection of $Q$ with $P(-37210300/12948573)$ is
\[
\bigg({-}\frac{9333678816679178366330705702}{22460621861042274504800702759}, -\frac{80704486762119296736852734835}{89842487444169098019202811036}\bigg),
\]
and the non-base point intersection of $Q$ with $P(-17169083242044/13796901723833)$ is
\begin{gather*}
\bigg(\frac{8837578153661106499492508487599309494380015}{3038431764392156856522470880096484622269351},
\\ \qquad
{}-\frac{67326893085562606742284626232543966149925005}{24307454115137254852179767040771876978154808}\bigg).
\end{gather*}
In Figures~\ref{f6}, the non-base point intersection of $B$ with $P(-37210300/12948573)$ is
\[
\bigg(\frac{3405617043949343809719957414870}{2300148977745403069873998022303}, \frac{19568184082895458339483642588681}{9200595910981612279495992089212}\bigg),
\]
and the non-base point intersection of $Q$ with $P(-17169083242044/13796901723833)$ is
\begin{gather*}
\bigg({-}\frac{1708121913130392164253959202492088885616083137916763}
{1128060630129279953146116160050550093652807049241987},
\\ \qquad \frac{23594087034110982474932797732767927894864716720290971}
{9024485041034239625168929280404400749222456393935896}\bigg).
\end{gather*}

\subsection*{Acknowledgements}
The author is grateful for useful and detailed comments made by the referees, in particular for the further simplification of the map $\gamma$, and the comment about infinitely near base points.

\pdfbookmark[1]{References}{ref}
\LastPageEnding

\end{document}